\documentstyle[12pt,epsf]{article}
\input epsf.tex

\begin{document}
\hoffset = -1.4truecm \voffset = -1.5truecm
\date{}

\newcommand{\beq}{\begin{equation}}
\newcommand{\eeq}{\end{equation}}
\newcommand{\beqa}{\begin{eqnarray}}
\newcommand{\eeqa}{\end{eqnarray}}
\newcommand{\beqar}{\begin{eqnarray*}}

\title{{\bf Lowest Order Constrained Variational Calculation of the Polarized
Nuclear Matter with the Modern $AV_{18}$ Potential}}

\author{{\bf G.H. Bordbar \footnote{Corresponding author} \footnote{E-mail :
Bordbar@physics.susc.ac.ir}} and {\bf M. Bigdeli}  \\
Department of Physics, Shiraz University,
Shiraz 71454, Iran\footnote{Permanent address}\\
and\\
Research Institute for Astronomy and Astrophysics of Maragha,\\
P.O. Box 55134-441, Maragha, Iran }


\maketitle


\begin{abstract}
The lowest order constrained variational method is applied to
calculate the polarized symmetrical nuclear matter properties with
the modern $AV_{18}$ potential performing microscopic
calculations. Results based on the consideration of magnetic
properties show no sign of phase transition to a ferromagnetic
phase.
\end{abstract}
21.65.-f, 26.60.-c, 64.70.-p

\newpage
\section{Introduction}
The properties of dense matter is a subject that theoretical
physicists have desired to study. The magnetic property of nucleon
matter is of special interest in nuclear and astrophysics which
can be related directly with magnetic source of pulsars,
 rapidly rotating neutron stars with strong surface magnetic fields
in the range of $10^{12} -10^{13}$ Gauss \cite{shap,paci,gold}.
The most interesting and stimulating mechanisms that have been
suggested is the possible existence of a phase transition to a
ferromagnetic state at densities corresponding to the
theoretically stable neutron stars and, therefore, of a
ferromagnetic core in the liquid interior of such compact objects.
Such a possibility has been studied by several authors using
different theoretical approaches [4-24], but the results are still
contradictory. Vidana et al. \cite{vida}, Vidana and Bombaci
\cite{vidab} have considered properties of spin polarized neutron
matter and polarized isospin asymmetric nuclear matter using the
Brueckner-Hartree-Fock (BHF) approximation by employing three
realistic nucleon- nucleon interactions, Nijmegen II, Reid93 and
NSC97e respectively. Zuo et al. \cite{zls} have also obtained
properties of spin polarized neutron and symmetric nuclear matter
using same method with $AV_{18}$ potential. The results of those
calculations show no indication of ferromagnetic transition at any
density for neutron and asymmetrical nuclear matter. Fantoni et
al. \cite{fanto} have calculated spin susceptibility of neutron
matter using the Auxiliary Field Diffusion Monte Carlo (AFDMC)
method employing the AU6 + UIX three-body potential, and have
found that the magnetic susceptibility of neutron matter shows a
strong reduction of about a factor 3 with respect to its Fermi gas
value. Baldo et al. \cite{bgls}, Akmal et al. \cite{apr} and
Engvik et al. \cite{ehmmp} have considered  properties of  neutron
matter with
 $AV_{18}$ potential using BHF approximation both for continuous choice
(BHFC) and standard choice (BHFG), variational chain summation
(VCS) method and lowest order Brueckner (LOB) respectively. On the
other hand some calculations, like for instance the ones based on
Skyrmelike interactions predict the transition to occur at
densities in the range $(1-4)\rho_{0}$ ($\rho_{0} = 0.16 fm^{-3}$)
\cite{apv}  . This transition could have important consequences
for the evolution of a protoneutron star, in particular for the
spin correlations in the medium which do strongly affect the
neutrino cross section and the neutrino mean free path inside the
star \cite{navarro}.

Recently, we have used the lowest order constrained variational
(LOCV) method \cite{owen} to calculate the equation of state of
symmetrical and asymmetrical nuclear matter and some of their
properties such as symmetry energy, pressure, etc. [32-35]. We
have also obtained the properties of spin polarized liquid
$^{3}He$ \cite{bord05} using this method. The LOCV method is a
useful tool for the determination of the properties of neutron,
nuclear and asymmetric nuclear matter at zero and finite
temperature. It is a fully self-consistent formalism which does
not bring any free parameters into calculation. It employs a
normalization constraint to keep the higher order term as small as
possible \cite{owen}. The functional minimization procedure
represents an enormous computational simplification over
unconstrained methods that attempt to go beyond lowest order.

In our pervious work, we have developed the LOCV method to compute
the properties of polarized neutron matter such as total energy,
magnetic susceptibility, pressure, etc. \cite{bordbig}, and have
seen that the spontaneous phase transition to a ferromagnetic
state in the neutron matter does not occur. In the present work,
we intend to calculate the polarized symmetrical nuclear matter
properties using the LOCV method with the modern $AV_{18}$
potential \cite{wiring} employing microscopic calculations.

\section{LOCV FORMALISM}
We consider a cluster expansion of the energy functional up to the
two-body term,
 \begin{eqnarray}
           E([f])=\frac{1}{A}\frac{\langle\psi|H|\psi\rangle}{\langle\psi|\psi\rangle}=E _{1}+E
           _{2}\ .
 \end{eqnarray}
The smallness of the three-body cluster energy has been discussed
in Ref. [32], where it is shown that our cluster expansion
converges reasonably and it is good approximation to stop after
the two-body energy term. This property can also be predicted by
looking at the correlation between the particles which will be
discussed in the next section.

The one-body term $E _{1}$  can be written as Fermi momentum
functional ($k _{F}^{i}=(3\pi^{2}\rho^{(i)})^{\frac{1}{3}})$,
\begin{eqnarray}\label{ener1}
               E _{1}=\sum _{i=1,2}\frac{3}{5}\frac{\hbar^{2}{k
               _{F}^{i}}^2}{2m}\frac{\rho^{(i)}}{\rho}\ .
 \end{eqnarray}
Labels 1 and 2 are used instead of spin up and spin down nucleons,
respectively, and $\rho=\rho^{1}+\rho^{2}$ is the total nuclear
matter density. The two-body energy $E_{2}$ is
\begin{eqnarray}
    E_{2}&=&\frac{1}{2A}\sum_{ij} \langle ij\left| \nu(12)\right|
    ij-ji\rangle,
 \end{eqnarray}
where
\begin{eqnarray}
\nu(12)=-\frac{\hbar^{2}}{2m}[f(12),[\nabla
_{12}^{2},f(12)]]+f(12)V(12)f(12)
\end{eqnarray}
$f(12)$ and $V(12)$ are the two-body correlation and potential,
respectively. For the two-body correlation function, $f(12)$, we
consider the following form \cite{borda,bordb}:
\begin{eqnarray}
f(12)&=&\sum^3_{k=1}f^{(k)}(12)O^{(k)}(12),
\end{eqnarray}
where, the operators $O^{(k)}(12)$ are given by
\begin{eqnarray}
O^{(k=1-3)}(12)&=&1,\ (\frac{2}{3}+\frac{1}{6}S_{12}),\
(\frac{1}{3}-\frac{1}{6}S_{12}),
\end{eqnarray}
and $S_{12}$ is the tensor operator. A complete discussion of
correlation function and especially its form are given in Ref.
\cite{owen}.

After doing some algebra, we find the following equation for the
two-body energy of the polarized symmetrical nuclear matter,
\begin{eqnarray}\label{ener2}
E_{2} &=& \frac{2}{\pi ^{4}\rho }\left( \frac{h^{2}}{2m}\right)
\sum_{JLTSS_{z}}\frac{(2J+1)(2T+1)}{2(2S+1)}[1-(-1)^{L+S+T}]\left|
\left\langle \frac{1}{2}\sigma _{z1}\frac{1}{2}\sigma _{z2}\mid
SS_{z}\right\rangle \right| ^{2} \nonumber \\&& \times \int
dr\left\{\left[{f_{\alpha }^{(1)^{^{\prime }}}}^{2}{a_{\alpha
}^{(1)}}^{2}(k_{f}r)+\frac{2m}{h^{2}}\left(\{V_{c}-3V_{\sigma }
+(V_{\tau }-3V_{\sigma \tau })(4T-3)
\right.\right.\right.\nonumber
\\&&\left.\left.\left.\ \ \ \ \ \ \ \ \ +(V_{T}-3V_{\sigma \tau })(4T)\}{a_{\alpha
}^{(1)}}^{2}(k_{f}r)\+[V_{l2}-3V_{l2\sigma }
\right.\right.\right.\nonumber
\\&&\left.\left.\left.\ \ \ \ \ \ \ \ \
+(V_{l2\tau}-3V_{l2\sigma \tau })(4T-3)]{c_{\alpha
}^{(1)}}^{2}(k_{f}r)\right)(f_{\alpha }^{(1)})^{2}\right]
 \right. \nonumber
\\&& \left.\ \ \ \ \ \ \ \ \ \ +\sum_{k=2,3}\left[ {f_{\alpha }^{(k)^{^{\prime }}}}^{2}{a_{\alpha
}^{(k)}}^{2}+\frac{2m}{h^{2}}\left(\left\{V_{c}+V_{\sigma
}+(-6k+14)V_{t}+-(k-1)V_{ls}\right.\right.\right.\right. \nonumber
\\&&\left.\left.\left.\left.\ \ \ \ \ \ \ \ \ \ \ \ \ \ \ \ \ +[V_{\tau } +V_{\sigma \tau
}+(-6k+14)V_{tz}-(k-1)V_{ls\tau
}](4T-3)\right.\right.\right.\right. \nonumber
\\&& \left.\left.\left.\left.\ \ \ \ \ \ \ \ \ \ \ \ \ \ \ \ \ +[V_{T}+V_{\sigma \tau
}+(-6k+14)V_{tT}] [4T]\right\}{a_{\alpha }^{(i)}}^{2}(k_{f}r)
\right.\right.\right.\nonumber
\\&&\left.\left.\left. \ \ \ \ \ \ \ \ \ \ \ \ \ \ \ \ \ +[V_{l2}+V_{l2\sigma } +(V_{l2\tau }+V_{l2\sigma \tau
})(4T-3)]{c_{\alpha
}^{(i)}}^{2}(k_{f}r)\right.\right.\right.\nonumber
\\&&\left.\left.\left.\ \ \ \ \ \ \ \ \ \ \ \ \ \ \ \ \ +[(V_{ls2}+V_{ls2\tau
})(4T-3)]{d_{\alpha }^{(k)}}^{2}(k_{f}r)\right) {f_{\alpha
}^{(k)}}^{2}\right] \right.\nonumber \\&&\left. \ \ \ \ \ \ \ \ \
\ \ \ \ \ \ \ \ \ +\frac{2m}{h^{2}}[[(V_{ls\tau}-2(V_{l2\sigma
\tau }+V_{l2\tau })-3V_{ls2\tau })(4T-3)]\right. \nonumber
\\&& \left.\ \ \ \ \ \ \ \ \ \ \ \ \ \ \ \ \ \ +V_{ls}-2(V_{l2}+V_{l2\sigma
})-3V_{ls2}]b_{\alpha }^{2}(k_{f}r)f_{\alpha }^{(2)}f_{\alpha
}^{(3)} \right. \nonumber
\\&& \left.\ \ \ \ \ \ \ \ \ \ \ \ \ \ \ \ \ \
+\frac{1}{r^{2}}(f_{\alpha }^{(2)} -f_{\alpha
}^{(3)})^{2}b_{\alpha }^{2}(k_{f}r)\right\}
 \end{eqnarray}
where $\alpha=\{J,L,S,S_z\}$ and the coefficient  ${a_{\alpha
}^{(1)}}^{2}$, etc. are defined as follows,
\begin{eqnarray}
     {a_{\alpha }^{(1)}}^{2}(x)=x^{2}I_{L,S_{z}}(x)
 \end{eqnarray}
\begin{eqnarray}
     {a_{\alpha }^{(2)}}^{2}(x)=x^{2}[\beta I_{J-1,S_{z}}(x)+\gamma I_{J+1,S_{z}}(x)]
 \end{eqnarray}
\begin{eqnarray}
           {a_{\alpha }^{(3)}}^{2}(x)=x^{2}[\gamma I_{J-1,S_{z}}(x)+\beta I_{J+1,S_{z}}(x)]
      \end{eqnarray}
\begin{eqnarray}
     b_{\alpha }^{(2)}(x)=x^{2}[\beta _{23}I_{J-1,S_{z}}(x)-\beta _{23}I_{J+1,S_{z}}(x)]
 \end{eqnarray}
\begin{eqnarray}
         {c_{\alpha }^{(1)}}^{2}(x)=x^{2}\nu _{1}I_{L,S_{z}}(x)
      \end{eqnarray}
\begin{eqnarray}
        {c_{\alpha }^{(2)}}^{2}(x)=x^{2}[\eta _{2}I_{J-1,S_{z}}(x)+\nu _{2}I_{J+1,S_{z}}(x)]
 \end{eqnarray}
\begin{eqnarray}
       {c_{\alpha }^{(3)}}^{2}(x)=x^{2}[\eta _{3}I_{J-1,S_{z}}(x)+\nu _{3}I_{J+1,S_{z}}(x)]
 \end{eqnarray}
\begin{eqnarray}
     {d_{\alpha }^{(2)}}^{2}(x)=x^{2}[\xi _{2}I_{J-1,S_{z}}(x)+\lambda _{2}I_{J+1,S_{z}}(x)]
 \end{eqnarray}
\begin{eqnarray}
     {d_{\alpha }^{(3)}}^{2}(x)=x^{2}[\xi _{3}I_{J-1,S_{z}}(x)+\lambda _{3}I_{J+1,S_{z}}(x)]
 \end{eqnarray}
with
\begin{eqnarray}
          \beta =\frac{J+1}{2J+1};\ \  \gamma =\frac{J}{2J+1}; \ \   \beta _{23}=\frac{2J(J+1)}{2J+1}
 \end{eqnarray}
\begin{eqnarray}
       \nu _{1}=L(L+1);\ \  \nu _{2}=\frac{J^{2}(J+1)}{2J+1};\ \  \nu _{3}=\frac{J^{3}+2J^{2}+3J+2}{2J+1}
      \end{eqnarray}
\begin{eqnarray}
     \eta _{2}=\frac{J(J^{2}+2J+1)}{2J+1}; \ \ \eta _{3}=\frac{J(J^{2}+J+2)}{2J+1}
 \end{eqnarray}
\begin{eqnarray}
     \xi _{2}=\frac{J^{3}+2J^{2}+2J+1}{2J+1};\ \   \xi _{3}=\frac{J(J^{2}+J+4)}{2J+1}
 \end{eqnarray}
\begin{eqnarray}
     \lambda _{2}=\frac{J(J^{2}+J+1)}{2J+1};\ \  \lambda _{3}=\frac{J^{3}+2J^{2}+5J+4}{2J+1}
 \end{eqnarray}

and
\begin{eqnarray}
       I_{J,S_{z}}(x)=\int dqP_{S_{z}}(q)J_{J}^{2}(xq)
 \end{eqnarray}
In the above equation, $J_{J}(x)$ is the Bessel's function and
$P_{S_{z}}(q)$ is defined as follows:
\begin{eqnarray}
       P_{S_{z}}(q)&=&\frac{2}{3}\pi \lbrack (k_{F}^{\sigma _{z1}})^{3}+(k_{F}^{\sigma _{z2}})^{3}
       -\frac{3}{2}((k_{F}^{\sigma _{z1}})^{2} +(k_{F}^{\sigma _{z2}})^{2})q
       \nonumber\\&&
       -\frac{3}{16}((k_{F}^{\sigma _{z1}})^{2}-(k_{F}^{\sigma _{z2}})^{2})^{2}q^{-1}+q^{3}]
 \end{eqnarray}
for $\frac{1}{2}\left| k_{F}^{\sigma _{z1}}-k_{F}^{\sigma
_{z2}}\right|
        < q< \frac{1}{2}\left| k_{F}^{\sigma _{z1}}+k_{F}^{\sigma
        _{z2}}\right|$,
\begin{eqnarray}
      P_{S_{z}}(q)=\frac{4}{3}\pi min(k_{F}^{\sigma _{z1}},k_{F}^{\sigma _{z2}})
 \end{eqnarray}
 for $q<\frac{1}{2}\left| k_{F}^{\sigma _{z1}}-k_{F}^{\sigma
 _{z2}}\right|$ and
 \begin{eqnarray}
            P_{S_{z}}(q)=0
 \end{eqnarray}
for $q>\frac{1}{2}\left| k_{F}^{\sigma _{z1}}+k_{F}^{\sigma
 _{z2}}\right|$,
where $\sigma _{z1}$ and $\sigma _{z2}$ are equal to
$\frac{1}{2},\ -\frac{1}{2}$ for spin up and spin down nucleons,
respectively.

Now, we minimize the two-body energy, Eq.(\ref{ener2}), with
respect to the variations in the correlation functions
${f_{\alpha}}^{(k)}$, but subject to the normalization constraint
\cite{bordb},
\begin{eqnarray}
        \frac{1}{A}\sum_{ij}\langle ij\left| h_{S_{z}}^{2}-f^{2}(12)\right| ij\rangle
        _{a}=0,
 \end{eqnarray}
where in the case of spin polarized nuclear matter, the function
$h_{S_{z}}(r)$ is defined as
\begin{eqnarray}
               h_{S_{z}}(r)&=& \left\{\begin{array}{ll}
              \left[ 1-\frac{9}{2}\left( \frac{J_{J}^{2}(k_{F}^{(i)}r)}{k_{F}^{(i)}r}\right) ^{2}\right] ^{-1/2} &~~ S_{z}=\pm1   \\ \\
              1 &~~ S_{z}= 0
                       \end{array}
                        \right.
            \end{eqnarray}

From the minimization of the two-body cluster energy, we get a set
of coupled and uncoupled
 differential equations which are the same as presented in Ref. \cite{bordb}.

\section{RESULTS}\label{NLmatchingFFtex}

In Fig. 1, we have shown the correlation function versus the
relative distance ($r$). Fig. 1 shows that the correlation between
particles is short range and heals to 1 very quickly. This means
that the two-body term mainly contributes to the interaction of
particles and therefore higher order terms can be neglected.

The energy per particle of the polarized symmetrical nuclear
matter versus density for different values of the spin
polarization have been shown in Fig. 2. This figure shows that the
low polarization gives more binding energy than the high
polarization. It is also seen that there is no crossing of the
energy curves of different polarizations, vice versa by increasing
density, the difference between the energy of nuclear matter at
different polarization becomes more sizable. This shows that the
spontaneous phase transition to a ferromagnetic state in the
symmetrical nuclear matter does not occur.

In Fig. 3, we have plotted the quadratic spin polarization
dependence  $\delta^{2}$ of energy per particle at different
densities. As can be seen from this figure, there are two points
worth stressing. First the energy per particle of the polarized
symmetrical nuclear matter increases as the polarization increases
and the minimum value of energy occurs at $\delta=0$ for all
densities. This indicates that the ground state of symmetrical
nuclear matter is paramagnetic. Second the variation of the energy
of symmetrical nuclear matter versus $\delta^{2}$ is nearly
linear,
\begin{eqnarray}
       E (\rho,\delta)&=&E(\rho,0)+a_{nucl}(\rho)\delta^{2}\cdot
 \end{eqnarray}
In Fig. 3, the results of ZLS calculations \cite{zls} are also
given for comparison. There is an agreement between our results
and those of ZLS, specially at low densities.

The magnetic susceptibility, $\chi$, which characterizes the
response of a system to the magnetic field and gives a measure of
the energy required to produce a net spin alignment in the
direction of the magnetic field, is defined by
\begin{eqnarray}\label{susc}
\chi =\left( \frac{\partial M}{\partial H}\right) _{H=0},
\end{eqnarray}
where $M$ is the magnetization of the system per unit volume and
$H$ is the magnetic field. We have calculated the magnetic
susceptibility of the polarized symmetrical nuclear matter in the
ratio ${{\chi}/{\chi_{F}}}$ form. By using the Eq. \ref{susc} and
some simplification, the ratio of $\chi$ to the magnetic
susceptibility for a degenerate free Fermi gas $\chi_{F}$ can be
written as
\begin{eqnarray}
   \frac{\chi}{\chi_{F}} =\frac{2}{3}\frac{E_{F}
}{\left( \frac{\partial ^{2}\left( E/N\right) }{\partial \delta
^{2}}\right) _{\delta =0}}\ ,
\end{eqnarray}
where $E_{F}={\hbar ^{2}k_{F}^{2}}/{2m}$ is the Fermi energy and
$k_{F}=(3/2\pi^{2}\rho)^{1/3}$ is Fermi momentum. Our results for
magnetic susceptibility are displayed  as a function of density in
Fig. 4. As can be seen from Fig. 4, this ratio changes
continuously for all densities and decreases as the density
increases. Therefore, the ferromagnetic phase transition is not
predicted by our calculation. For comparison, we have also shown
the results of ZLS \cite{zls} in this figure which shows good
agrement whit our results.

By differentiating symmetrical nuclear matter energy curve at each
polarization ($\delta$) whit respect to the density we can
evaluate the corresponding pressure,
\begin{eqnarray}
      P(\rho,\delta)= \rho^{2} \frac{\partial {{E}}(\rho,\delta)}{\partial
      \rho}\ ,
 \end{eqnarray}
In Fig. 5, we have shown the pressure of polarized symmetrical
nuclear matter as a function of density $\rho$ for various
polarizations. We see that equation of state of polarized
symmetrical nuclear matter, $P(\rho,\delta)$, becomes stiffer by
increasing the polarization in the density range which was
considered.

In Fig. 6, we have also presented the Landau parameter, $G_{0}$,
which describes the spin density fluctuation in the effective
interaction, versus density. It is seen that the value of $G_{0}$
is always positive and monotonically increasing up to highest
density and does not show any magnetic instability for the neutron
matter. A magnetic instability would require $G_{0}<-1$.

\section{Summary and Conclusions}

We have computed the magnetic properties of polarized symmetrical
nuclear matter, that is related directly with magnetic source of
pulsars, and other properties using the lowest order constrained
variational (LOCV) method with with $AV_{18}$ potential. We have
studied the total energy per particle of nuclear matter as a
function of density and the spin polarizations $\delta$. We have
found that in the range of densities explored, difference between
the  energy of polarized nuclear matter at different polarization
becomes more appreciable. We have also seen that total energy per
particle is parabolic on the spin polarization $\delta$ in a very
good approximation up to full polarization for all densities.
Magnetic susceptibility, which characterizes the response of the
system to the magnetic field was calculated for the system under
consideration and was found that it changes continuously for all
densities. There is an overall agreement between our result and
those of Zuo et al. \cite{zls}. In conclusion, we see that
equation of state of polarized symmetrical nuclear matter becomes
stiffer by increasing the polarization in the density range which
was considered. The Landau parameter, $G_{0}$ has been considered
and it was seen that the value of $G_{0}$ is always positive and
monotonically increasing up to high densities. Finally, our
results have shown no phase transition to ferromagnetic state.

\section*{Acknowledgements}
This work has been supported by Research Institute for Astronomy
and Astrophysics of Maragha, and Shiraz University Research
Council.

\newpage

\newpage

\begin{figure}
\centerline{\epsfxsize 4.5 truein \epsfbox {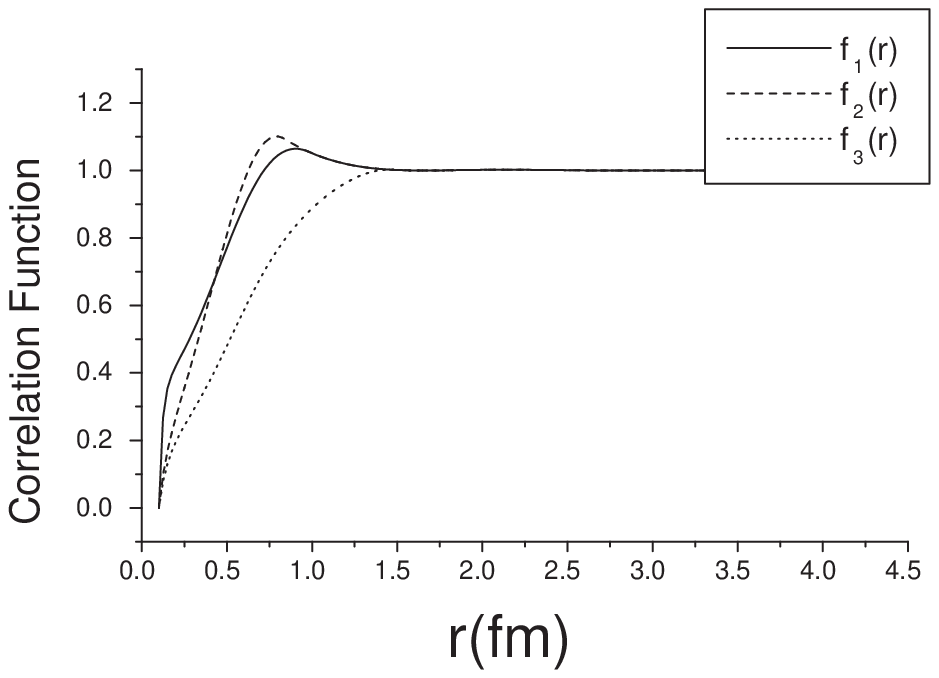}}
\caption{The two-body correlation functions of the full polarized
nuclear matter as a function of relative distance at $\rho = 0.67
fm^{-3} $.} \label{correlate)}
\end{figure}

\begin{figure}
\centerline{\epsfxsize 4.5 truein \epsfbox {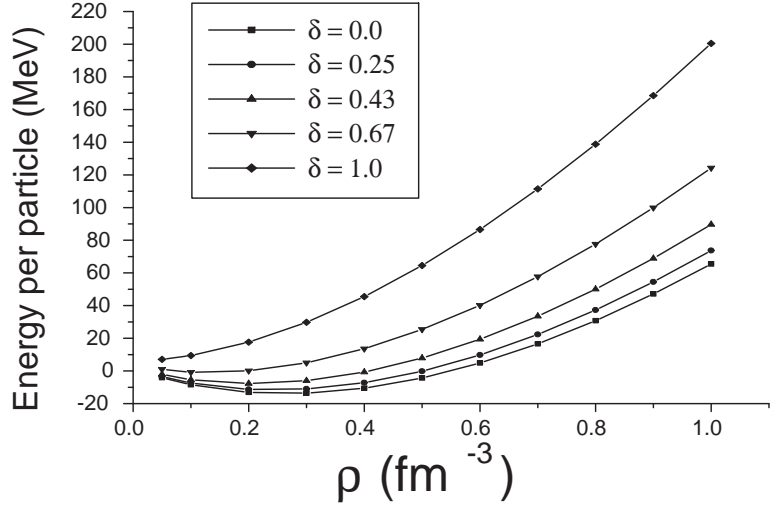}}
\caption{The energy per particle of the polarized symmetrical
nuclear matter versus density($\rho$) for different values of the
spin polarization ($\delta$).} \label{EN(den)}
\end{figure}

\begin{figure}
\centerline{\epsfxsize 4.5 truein \epsfbox {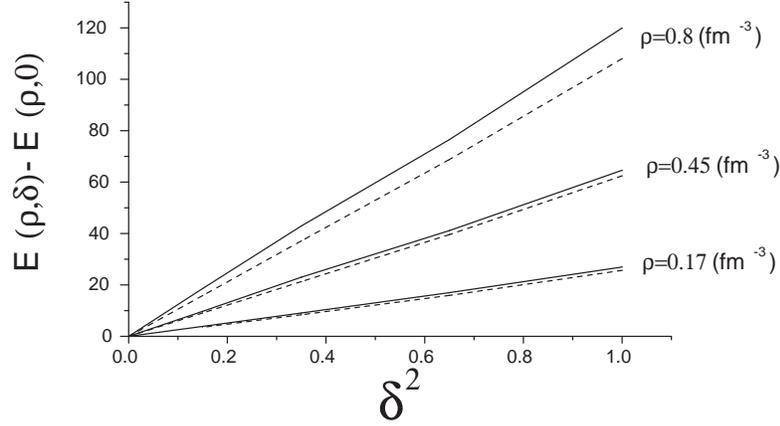}}
\caption{Our results (full curves) for the energy difference of
polarized and unpolarized cases versus quadratic spin polarization
($\delta^{2}$) for different values of the density($\rho$) of the
neutron matter. The results of ZLS [25] (dashed curves) are also
presented for comparison.} \label{EN(comp)}
\end{figure}
\newpage


\begin{figure} \centerline{\epsfxsize 4.5 truein \epsfbox
{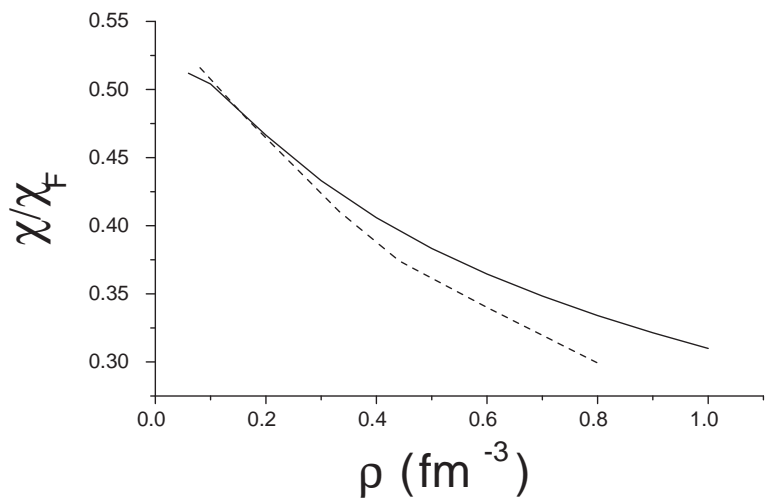}} \caption{Our result(full curve) for the magnetic
susceptibility of the polarized symmetrical nuclear matter as the
function of density($\rho$). The results of ZLS [25] (dashed
curves) are also given for comparison.} \label{sus(den)}
\end{figure}


\begin{figure}
\centerline{\epsfxsize 4.5 truein \epsfbox {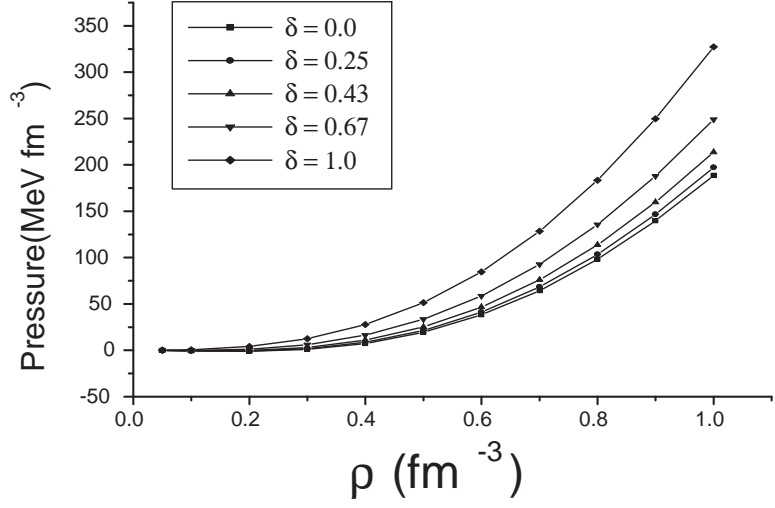}}
\caption{The equation of state of polarized symmetrical nuclear
matter for different values of the spin polarization ($\delta$).}
\label{PRE(den)}
\end{figure}



\begin{figure}
\centerline{\epsfxsize 4.5 truein \epsfbox {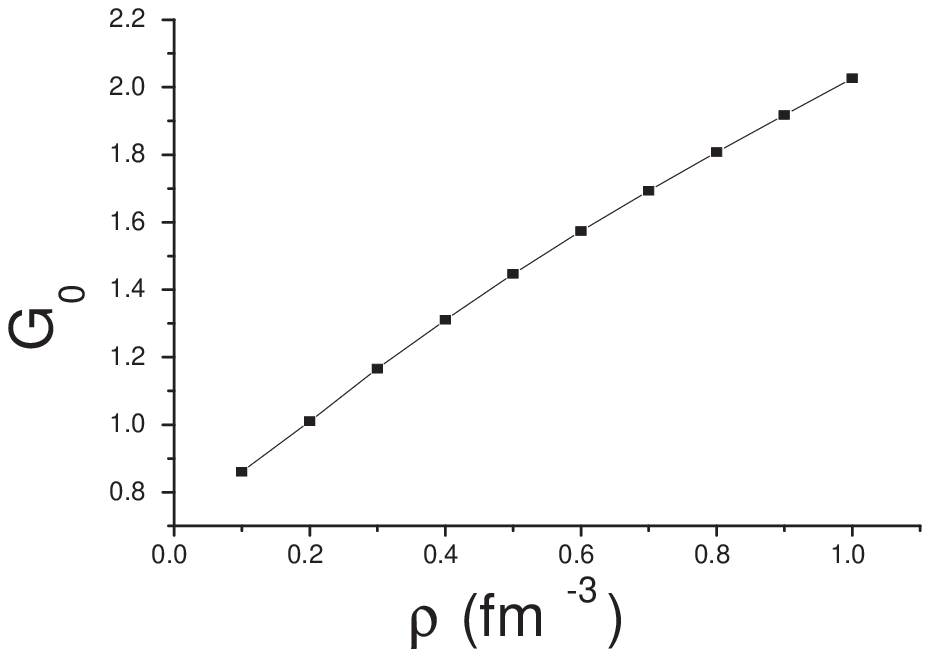}}
\caption{Our result for the Landau parameter, $G_{0}$, as function
of density($\rho$).} \label{G(den)}
\end{figure}



\begin{thebibliography}{99}
\bibitem{shap} S. Shapiro and S. Teukolsky, Blak Holes, White Dwarfs and Neutron Stars, (Wiley-New york,1983).
\bibitem{paci} F. Pacini, \emph{Nature} (London) {\bf 216} (1967) 567.
\bibitem{gold} T. Gold, \emph{Nature} (London) {\bf 218} (1968) 731.
\bibitem{brown} D. H. Brownell and J. Callaway, \emph{Nuovo Cimento} {\bf B 60} (1969) 169.
\bibitem{rice} M. J. Rice, \emph{Phys. Lett}. {\bf A 29} (1969) 637.
\bibitem{clark} J. W. Clark and N. C. Chao, \emph{Lettere Nuovo Cimento} {\bf 2} (1969) 185.
\bibitem{clark2} J. W. Clark, \emph{Phys. Rev. Lett}. {\bf 23} (1969) 1463.
\bibitem{silv} S. D. Silverstein, \emph{Phys. Rev}. Lett. {\bf 23} (1969) 139.
\bibitem{Østga} E. {\O}stgaard, \emph{Nucl. Phys}. {\bf A 154} (1970) 202.
\bibitem{pear} J. M. Pearson and G. Saunier, \emph{Phys. Rev. Lett}. {\bf 24} (1970) 325.
\bibitem{pandh} V. R. Pandharipande, V. K. Garde and J. K. Srivastava, \emph{Phys. Lett}. {\bf B 38} (1972) 485.
\bibitem{backm}  S. O. Backman and C. G. Kallman, \emph{Phys. Lett}. {\bf B 43} (1973) 263.
\bibitem{haens} P. Haensel, \emph{Phys. Rev}. {\bf C 11} (1975) 1822.
\bibitem{jack}  A. D. Jackson, E. Krotscheck, D. E. Meltzer and R. A. Smith, \emph{Nucl. Phys}. {\bf A
386} (1982)125.
\bibitem{kuts}   M. Kutschera and W. W´ojcik, \emph{Phys. Lett}. {\bf B 223} (1989) 11.
\bibitem{marcos} S. Marcos, R. Niembro, M. L. Quelle and J. Navarro, \emph{Phys. Lett}. {\bf B 271} (1991) 277.
\bibitem{bern} P. Bernardos, S. Marcos, R. Niembro, M. L. Quelle, \emph{Phys. Lett}. {\bf B
356} (1995) 175.
\bibitem{vidau} A. Vidaurre, J. Navarro and J. Bernabeu, \emph{Astron. Astrophys}. {\bf
135} (1984) 361.
\bibitem{kutsb} M. Kutschera and W. W´ojcik, \emph{Phys. Lett}. {\bf B 325} (1994) 271.
\bibitem{fanto}S. Fantoni, A. Sarsa and K. E. Schmidt, \emph{Phys. Rev. Lett}. {\bf 87} (2001) 181101.
\bibitem{vida}I. Vida˜na, A. Polls and A. Ramos, \emph{Phys. Rev}. {\bf C 65} (2002) 035804.
\bibitem{vidab} I. Vida˜na and I. Bombaci, \emph{Phys. Rev}. {\bf C 66} (2002) 045801.
\bibitem{zuo}W. Zuo, U. Lombardo and C.W. Shen, in Quark-Gluon Plasma and
Heavy Ion Collisions, Ed. W.M. Alberico, M. Nardi and M.P.
Lombardo, World Scientific, p. 192 (2002).
\bibitem{isay}A. A. Isayev and J. Yang, \emph{Phys. Rev}. {\bf C 69} (2004) 025801.
\bibitem{zls} W. Zuo, U. Lombardo and C. W. Shen,
nucl-th/0204056.\\
W. Zuo, C. W. Shen and U. Lombardo, Phys. Rev {\bf C 67} (2003)
037301.
\bibitem{bgls} M. Baldo, G. Giansiracusa, U. Lombardo and H. Q.
Song, Phys. Lett. {\bf B 473} (2000) 1.
\bibitem{apr} A. Akmal, V. R. Pandharipande and D. G. Ravenhall,
Phys. Rev. {\bf C 58} (1998) 1804.
\bibitem{ehmmp} L. Engvik et al., Nucl. Phys. {\bf A 627} (1997)
85.
\bibitem{apv} A.Rios, A. Polls and I. Vidana, Phys. Rev. {\bf C
71}(2005) 055802.
\bibitem{navarro}J. Navarro, E. S. Hern´andez and D. Vautherin,
\emph{Phys. Rev}. {\bf C 60} (1999) 045801.
\bibitem{owen}J. C. Owen, R. F. Bishop, and J. M. Irvine,
\emph{Nucl. Phys.} {\bf A 277} (1977) 45.
\bibitem{borda}G. H. Bordbar, M. Modarres, \emph{J. Phys. G: Nucl. Part. Phys}.
{\bf 23} (1997) 1631.
\bibitem{bordb}G. H. Bordbar and M. Modarres, \emph{Phys. Rev.} {\bf C
57} (1998) 714.
\bibitem{mbord98}M. Modarres and G. H. Bordbar, \emph{Phys. Rev.} {\bf C
58} (1998) 2781.
\bibitem{bord03}G. H. Bordbar, \emph{Int. J. M. Phys.} {\bf A 18}
(2003) 3629.
\bibitem{bord05}G. H. Bordbar and S. M. Zebarjad, M. R. Vahdani, M.
Bigdeli, \emph{Int. J. M. Phys.} {\bf B 23} (2005) 3379.
\bibitem{bordbig}G. H. Bordbar and M. Bigdeli, \emph{Phys. Rev.}
{\bf C 75} (2007) 045804.
\bibitem{wiring}R. B. Wiringa, V. Stoks and R. Schiavilla,
\emph{Phys. Rev.} {\bf C
75} (1995) 38.
\end{thebibliography}
\end{document}